\documentclass[twocolumn,nofootinbib]{revtex4-1}
\usepackage{epsfig}

              \newcommand{\rf}[1]{(\ref{#1})}

\def\bfone{\relax{\rm 1\kern-.35em 1}}

\newcommand{\be}{\begin{equation}}
\newcommand{\ee}{\end{equation}}
\newcommand{\ben}{\begin{displaymath}}
\newcommand{\een}{\end{displaymath}}
\newcommand{\bea}{\begin{eqnarray}}
\newcommand{\eea}{\end{eqnarray}}

\newcommand{\bean}{\begin{eqnarray*}}
\newcommand{\eean}{\end{eqnarray*}}

\newcommand{\vp}{\varphi}

\def\K{K{\"a}hler}

\begin{document}

\title{\Large{Single-field $\alpha$-attractors}}

\author{Andrei Linde}

\affiliation{Department of Physics and SITP, Stanford University, \\ 
Stanford, California 94305 USA, alinde@stanford.edu}

\begin{abstract}
I describe a simple class of $\alpha$-attractors, generalizing the single-field GL model of inflation in supergravity. The new class of models is defined for $0<\alpha \lesssim 1$, providing a good match to the present cosmological data.  I also present a generalized version of these models which can describe not only inflation but also dark energy and supersymmetry breaking.
\end{abstract}

\maketitle

\smallskip


\section{Introduction}\label{intro}


First models of inflation in  supergravity  were based on the new inflationary scenario, assuming high temperature phase transitions with symmetry restoration. But these models did not quite work, and in 1983 the new inflation scenario was dethroned by chaotic inflation   \cite{Linde:1983gd}. 

The main idea of chaotic inflation was to consider various sufficiently flat potentials, either large-field or small-field, and check whether inflation may occur in some parts of the universe without assuming that it was in a state of thermal equilibrium and that initial state of the inflaton field should correspond to an extremum of the potential. For several years, this simple idea was rejected by many as a drastic deviation from the main principles of inflation, but gradually it became broadly accepted, and now practically all  inflationary models are based on it.

The first model of chaotic inflation in supergravity was proposed in 1983-1984 \cite{Goncharov:1983mw}; I will call it GL model hereafter. It was also the first model with the inflaton potential asymptotically approaching a plateau, $V   \sim a- b e^{-c\phi}$. Later on, it was realized that the Starobinsky model \cite{Starobinsky:1980te}, after certain modifications, can be cast in a form with a similar plateau potential \cite{Whitt:1984pd}. 

Inflationary potentials in these models never reach Planckian values. It took many years to solve the problem of initial conditions there, see a discussion in \cite{Linde:2014nna}. These models attracted general attention only recently, because they were strongly favored by the WMAP and Planck data \cite{Hinshaw:2012aka,Planck:2015xua}.  Predictions of some of these models are stable with respect to even large changes of their potentials;  such models are called cosmological attractors \cite{Kallosh:2013hoa,Kallosh:2013tua,Ferrara:2013rsa,Kallosh:2013yoa,Cecotti:2014ipa,Galante:2014ifa,Kallosh:2015zsa}. In particular, GL model belongs to the class of $\alpha$-attractors \cite{Ferrara:2013rsa,Kallosh:2013yoa,Cecotti:2014ipa} with $\alpha = 1/9$ \cite{Linde:2014hfa,Kallosh:2015lwa}. These models have a unique set of predictions  providing an excellent fit to the recent observational data for $\alpha \lesssim O(30)$:
\be\label{nsr}
 1 -n_{s} \approx {2\over N}\, , \qquad r \approx  {12\alpha \over N^{2} } \ .
 \ee
GL model \cite{Goncharov:1983mw} has several different realizations. It can be represented 
as a theory with a canonical \K\ potential 
\be\label{shift}
K = -{1\over 2} (\Phi-\bar\Phi)^{2}
\ee
and a superpotential \cite{Linde:2014hfa}
\be\label{cosh}
W = {m\over 6}\, \bigl(\cosh{\sqrt{3}\Phi} - \cosh^{-1}{\sqrt{3}\Phi}\bigr)\ .
\ee

From the point of view of the theory of $\alpha$-attractors, it is more appropriate to use logarithmic \K\ potentials, such as \cite{Cecotti:2014ipa},
\be\label{ka}
K= -3  \log \Big (1- Z\bar Z  + {\alpha-1\over 2} {(Z-\bar Z)^2\over 1- Z\bar Z}\Big )\,  
\ee
with $\alpha = 1/9$. In this framework, the GL model has a very simple superpotential \cite{Kallosh:2015lwa}
\be\label{GL}
 W =  {\mu\over 9}\, Z^{2}\, (1-Z^{2})  \ .
\ee
The inflationary potential of this model, upon transformation to the canonically normalized inflaton field  $\vp$ such that ${\rm Re}\, Z = \tanh{\vp\over\sqrt{6\alpha}}$, becomes
\be\label{potG}
V(\phi)= {\mu^2\over 27}  \Bigl(4 -  \tanh^{ 2}\sqrt{3\over 2} \vp\Bigr)\,  \tanh^{ 2}\sqrt{3\over 2} \vp\ \ .
\ee
It has a minimum at $\vp = 0$, where it vanishes. At $\vp \gtrsim 1$, the potential coincides with the plateau potential
\be\label{appG}
V(\vp) = {\mu^{2}\over 9} \left(1- {8\over 3}\, e^{-\sqrt 6 |\vp|}\right) \ ,
\ee 
up to exponentially small higher order corrections  \cite{Goncharov:1983mw}.

This model is quite economical: It involves just a single chiral superfield. It is very difficult to construct such models, so 
most of the subsequently developed inflationary models in supergravity involved at least two different supermultiplets. 

This situation changed only very recently, with invention of some interesting single superfield inflation models \cite{Ferrara:2013rsa,Ketov:2014qha,Linde:2014ela}, and especially with the development of models with nilpotent chiral superfields, which allow to have two superfields but only one complex scalar field  \cite{Ferrara:2014kva,Kallosh:2014via,Dall'Agata:2014oka,Kallosh:2014hxa,Lahanas:2015jwa,Kallosh:2015lwa,prep}. There are several different ways to incorporate inflationary models with any value of $\alpha$ in the context of such theories, and simultaneously describe a non-zero cosmological constant and SUSY breaking \cite{Kallosh:2014hxa,Kallosh:2015lwa}. 

In fact, even the original GL model, as well as the models of Ref. \cite{Ketov:2014qha,Linde:2014ela}, require  additional fields to describe SUSY breaking and the cosmological constant, but  one can easily achieve it by adding a tiny superpotential  $M(S+1/b)$ of a nilpotent field $S$ to the original single-filed GL superpotential \cite{Linde:2014hfa,Kallosh:2015lwa}. 
Therefore it would be interesting to find other examples of single-field models of this type which could incorporate various values of $\alpha$, not just $\alpha = 1/9$, and to check whether one could generalize them in a similar way.

The first part of this challenging problem was recently solved by Roest and Scalisi \cite{Roest:2015qya}. The authors studied models with \K\ potentials \cite{Kallosh:2013yoa}
\be\label{alphaK}
K = -3\alpha \log(T+\bar T) 
\ee
and found a family of $\alpha$-attractors with superpotentials
\be\label{DMP}
W=T^{3(\alpha + \sqrt{\alpha})\over 2} \left(1 + 3\sqrt \alpha - 3 \sqrt\alpha\, T -T^{-3 \sqrt{\alpha}}\right) \ .
\ee
Here, following \cite{Roest:2015qya}, we ignored the normalization coefficient in front of $W$. Some generalizations of this superpotential are possible, such as
\be\label{DM}
 W=T^{3(\alpha+\sqrt\alpha)\over 2} \left({T^{-3\sqrt\alpha} -1\over 3\sqrt{\alpha}} +T-1\right)    \, .
 \ee 

As pointed out in \cite{Roest:2015qya}, these models describe stable inflationary behavior at ${\rm Im} \ T = 0$ for all sufficiently large $\alpha > 1$, which is a significant achievement. However, the inflationary trajectory ${\rm Im} \ T = 0$ in the models \rf{DMP} and \rf{DM} is unstable for $\alpha\leq 1$, and therefore they  represent inflationary $\alpha$-attractors only for $\alpha > 1$. 

For example, one could try to develop a new implementation of the GL model with $\alpha = 1/9$ in this context \cite{Roest:2015qya}, using the superpotential
\be\label{DSGT}
W = T^{-1/3}\,  {(1 -T)^2 \over 1 + T} \ .
\ee
However, whereas the potential of the inflaton field ${\rm Re} \ T $ in this model coincides with the GL inflaton potential, the inflationary trajectory ${\rm Im} \ T = 0$ in this model is unstable, unlike the inflationary trajectory in the original GL model. 

Fortunately, one can find stable generalizations of the GL model for all positive $\alpha$ in the models with canonical \K\ potentials \rf{shift} \cite{Roest:2015qya}. This is  a very strong result, but the attractor nature of the models, which is related to the pole in the kinetic term of the inflaton field \cite{Galante:2014ifa}, is manifest only in the models with logarithmic \K\ potentials such as \rf{ka} or \rf{alphaK}. 

Since we already have a family of $\alpha$-attractors with with logarithmic \K\ potentials and nilpotent fields, covering the full range of $\alpha$ and containing only one dynamical scalar degree of freedom \cite{Dall'Agata:2014oka,Kallosh:2014hxa,Lahanas:2015jwa,Kallosh:2015lwa,prep}, it would be interesting to know whether one can achieve a similar success in the theory  with a single superfield.

In this short note we take a next step in this direction and 
present a complementary class of single-field inflationary $\alpha$-attractors, which are stable for $\alpha \lesssim 1$. We will also generalize these models to allow for a controllable level of supersymmetry breaking and vacuum energy. 

The new  $\alpha$-attractors, which we introduce here, have the \K\ potential \rf{ka} and the superpotential
\bea\label{Z}
 W(Z)&=&(1-Z^{2})^{3(1-\sqrt\alpha)\over 2} \Big((1+Z)^{3\sqrt\alpha} \nonumber \\ &-& (1+6\sqrt \alpha\, Z)(1-Z)^{3\sqrt\alpha}\Big)    \, .
 \eea
 
 \newpage

For $\alpha = 1/9$ this theory  exactly reproduces the GL model in its latest formulation with the simple superpotential  $W =  {\mu\over 9}\, Z^{2}\, (1-Z^{2})$  \cite{Kallosh:2015lwa}. One can show that these models describe a stable inflationary trajectory with ${\rm Im} \ Z = 0$ and lead to the usual $\alpha$-attractor predictions  \rf{nsr} for all values of $\alpha$ in the range $0<\alpha <0.989$. (A possible exception involves very small $\alpha \ll 1/9$, where the inflaton field after inflation may overshoot the minimum at $Z = 0$.) Various generalizations of this model are possible, e.g. one can add a small quadratic term $c Z^{2}$ to $(1+6\sqrt \alpha\, Z)$ in \rf{Z}; see also \cite{Kallosh:2015lwa}.

SUSY breaking and dS uplifting in these models can be achieved following \cite{Linde:2014hfa,Kallosh:2015lwa}. One can introduce a nilpotent superfield $S$, which does not have any scalar degrees of freedom, and use the \K\ potential
\be
\label{ka1}
K= -3  \log \Big (1- Z\bar Z  + {\alpha-1\over 2} {(Z-\bar Z)^2\over 1- Z\bar Z} -{S\bar S\over 3}\Big)\, . 
\ee
The SUSY breaking superpotential can be taken as a sum of $W(Z)$ \rf{Z} and a simple Polonyi-type superpotential of the nilpotent field $S$:
\be
W(Z,S) = W(Z) +M(S+1/b) 
\ee
with $M\ll 1$. The last term can be neglected during inflation, but it provides the required uplifting to a dS vacuum with a small cosmological constant. The minimum of the potential remains at $\vp =0$,  supersymmetry is spontaneously broken at the minimum,
\be
D_S W =  M\ , \quad  D_Z W = 0\ , \quad m_{3/2}=  {M/b}\ .
\label{dSpar}\ee
Vacuum energy generically is non-zero, 
\be
V_{0}=   {M^{2} (1-3/b^{2}}) \ .
\label{dSpar2}\ee
Note that $V_{0}$ is proportional to $M^{2}$, so dS uplifting is possible only because supersymmetry is spontaneously broken \cite{Linde:2014hfa,Kallosh:2015lwa}. By tuning $b\approx
\sqrt 3$ one can achieve any value of the cosmological constant, including the desirable value $V_{0}~\sim~10^{{-120}}$, along the lines of the string landscape scenario.

Thus in this new class of single-field $\alpha$-attractors one can simultaneously describe inflation, dark energy/cosmological constant, and SUSY breaking of a controllable magnitude, for all $\alpha \lesssim 1$.  This complements the results of \cite{Roest:2015qya} describing single field $\alpha$-attractors with a logarithmic \K\ potential which are stable for $\alpha > 1$. It would be nice to find a similar mechanism of SUSY breaking and uplifting for $\alpha$-attractors introduced in \cite{Roest:2015qya}, and   to find a way to close the small gap between the two families of attractors by stabilizing both sets of models in the vicinity of  $\alpha = 1$.

I am grateful to D. Roest and M. Scalisi for the continuing collaboration and for sharing with me the results of \cite{Roest:2015qya}  prior to publication. I am supported by the SITP and by the NSF Grant PHY-1316699  and by the Templeton foundation grant `Inflation, the Multiverse, and Holography.'

\end{document}